**High Mobility in a Stable Transparent Perovskite Oxide**

Hyung Joon Kim[*], Useong Kim[1,*], Hoon Min Kim[1], Tai Hoon Kim, Hyo Sik Mun[1], Byung-Gu Jeon, Kwang Taek Hong, Woong-Jhae Lee, Chanjong Ju[1], Kee Hoon Kim[†], and Kookrin Char[1,‡]

*Center for Novel States of Complex Materials Research, Department of Physics and Astronomy, Seoul National University, Seoul 151-747, Republic of Korea*

[1]*Center for Strongly Correlated Materials Research, Department of Physics and Astronomy, Seoul National University, Seoul 151-747, Republic of Korea*

**Abstract**

**We discovered that La-doped $BaSnO_3$ with the perovskite structure has an unprecedentedly high mobility at room temperature while retaining its optical transparency. In single crystals, the mobility reached 320 $cm^2(Vs)^{-1}$ at a doping level of $8\times10^{19}$ $cm^{-3}$, constituting the highest value among wide-band-gap semiconductors. In epitaxial films, the maximum mobility was 70 $cm^2(Vs)^{-1}$ at a doping level of $4.4\times10^{20}$ $cm^{-3}$. We also show that resistance of $(Ba,La)SnO_3$ changes little even after a thermal cycle to 530 °C in air, pointing to an unusual stability of oxygen atoms and great potential for realizing transparent high-frequency, high-power functional devices.**

[*]These authors equally contributed to this work.
[‡]Co-corresponding author
[†]E-mail address: khkim@phya.snu.ac.kr

Recently, there has been a surge of interest in transparent electronic materials in an effort to combine display devices with other electronic functions such as a touch screen, providing both optical transparency and good electrical conduction[1,2]. The interest in transparent conducting oxides (TCOs) has been extended to an area of transparent oxide semiconductors (TOSs) with a desire to incorporate more active functions[3,4]. One of the major efforts in this area has been on ZnO-based materials, successfully demonstrating *pn* junctions[3], field-effect transistors[3], and lasers in UV spectra[4]. By carefully controlling the strain at the interface of ZnO/Zn$_{1-x}$Mg$_x$O heterostructure, successful two-dimensional electron gas structures were fabricated and their quantum Hall effects were observed[5]. An alternative metal oxide system that has a wide band gap and can be easily doped with carriers of high mobility will contribute significantly to this rapidly developing research field on TCOs and TOSs.

A variety of perovskite metal oxides with a primary structure of ABO$_3$ were studied to show diverse properties such as superconductivity[6] and ferroelectricity[7]. The perovskite structure offers an advantage of doping possibility in the two different cation sites. In this paper, we introduce such a perovskite oxide semiconductor BaSnO$_3$ [Fig. 1(a)] that has a wide band gap of 3.1 eV[8]. BaSnO$_3$ has been used as a high-temperature gas sensor[9] as well as nanoparticle pinning centers for high temperature superconductors[10]. Recently, it has been investigated as epitaxial films for its potential as TCOs but the films, either when doped with Sb in place of Sn or La in place of Ba, exhibited relatively low mobility values[11,12], no more than 2 cm$^2$(Vs)$^{-1}$. In contrast to the previous results, herein we show that BaSnO$_3$ indeed exhibits high mobility when its Ba site is doped with La in both single crystals and epitaxial films. More importantly, we find that the (Ba,La)SnO$_3$ maintains its transport properties at temperatures as high as 530 °C, pointing to extremely stable oxygen atoms.

Polycrystalline (Ba,La)SnO$_3$ was synthesized by solid-state reaction, using a stoichiometric mixture of high-purity BaCO$_3$, SnO$_2$, and La$_2$O$_3$ powders. The mixture was first calcined at 1250 $^o$C for 6 h and finally fired at 1400-1450 $^o$C for 24 h after several intermediate grindings. (Ba,La)SnO$_3$ single crystals were grown by the flux method in air, using a Pt crucible, Cu$_2$O flux, and the sintered powders as a seed material. Based on the electron probe microanalyses, we confirmed that Cu impurities do not exist in the grown crystals. Thin films were deposited by the pulsed laser deposition technique, using BaSnO$_3$, Ba$_{0.96}$La$_{0.04}$SnO$_3$, and Ba$_{0.93}$La$_{0.07}$SnO$_3$ targets. Thin films at a smaller La dooping rate than 4 at% were made by sequential deposition of the BaSnO$_3$ and Ba$_{0.96}$La$_{0.04}$SnO$_3$ targets. Perovskite SrTiO$_3$ (001) substrates were used and the deposition was made in an O$_2$ pressure of 0.1 Torr at 750 $^o$C. After the deposition, the samples were cooled in an O$_2$ pressure of about 0.6 Torr. Temperature-dependent resistivity was investigated by the conventional four-probe method. We also used the five-wire configuration to investigate the Hall effect in a magnetic field up to 6 Tesla and at the excitation current about 100$\mu$A. The van der Pauw method in a magnetic field of 0.52 Tesla was also employed to investigate the Hall effects of thin films at room temperature.

The grown single crystals of BaSnO$_3$ and (Ba,La)SnO$_3$ are about 1-2 mm in size and maintain optical transparency [Fig. 1(b)]. According to the X-ray diffraction (XRD) study, both single crystals and epitaxial thin films exhibited well-defined (00$l$) diffraction peaks [Fig. 1(c)]. Even in the 7 at% La-doped thin film, we found no evidence of a secondary phase, such as La$_2$Sn$_2$O$_7$, which is commonly found in ceramic samples[13]. This is probably because the growth temperature is lower for thin films and more dopants can be accommodated. Moreover, the lattice constants of the 200 nm thick $Ba_{0.96}La_{0.04}SnO_3$ film, measured from the reciprocal space mapping of (103) peaks [Fig. 1(d)], turned out to be 4.127 Å in the growth direction and 4.107 Å on the plane. This observation implies that there exists a slight compressive strain in our films in light of bulk cubic lattice constant of 4.116 Å as measured from the XRD data of our $BaSnO_3$ single crystals.

Fig. 2 summarizes the resistivity ($\rho$) and mobility ($\mu$) values measured for all our single crystals and thin films as a function of their carrier densities ($n$) at 300 K. The highest $\mu$ = 320 $cm^2(Vs)^{-1}$ was found in the lowest doped single crystal, which has $n = 8.0 \times 10^{19}$ $cm^{-3}$. We note that $\mu$ corresponds to the highest record among the wide-band-gap semiconductors with a similar doping level. For example, the GaN[14] has $\mu$ = ~100 $cm^2(Vs)^{-1}$, indium oxide ($In_2O_3$)[15] has $\mu$ = ~160 $cm^2(Vs)^{-1}$, tin oxide($SnO_2$)[16] has $\mu$ = ~50 $cm^2(Vs)^{-1}$, and ZnO[17] has $\mu$ = ~100 $cm^2(Vs)^{-1}$. The unusually high $\mu$ may result from the small effective mass of $BaSnO_3$ related to the ideal Sn-O-Sn bond angle (~180°). For example, the bandwidth and dispersion of the $BaSnO_3$ conduction band[8] are quite similar to those of $In_2O_3$[18], which guarantees a small effective mass ($m_{eff}$). It is noteworthy that recently reported theoretical $m_{eff}$ of $BaSnO_3$[19], $0.06m_0$, is much smaller than theoretical $m_{eff}$ of $SnO_2$ ($0.38m_0$)[20]. The high $\mu$ can be also associated with reduced dopant scattering due to rather a high dielectric constant (~20)[21], low disorder effect due to the Ba-site doping of La ions away from $SnO_2$ layers, and/or possibly small phonon scattering. Further studies will be needed to pin down the exact origin for the high $\mu$.

In Fig. 2, it is clear that the single crystals and thin films follow two completely different trends; $\mu$ of single crystals increases as $n$ decreases, while $\mu$ of thin films decreases as $n$ decreases in a low doping range smaller than $n = 4\times10^{20}/cm^3$. However, $\mu$ of thin films almost approaches the value of a single crystal around $n = 4\times10^{20}/cm^3$, and then eventually follows the single crystal trend at higher $n$, e.g., $n = 6.8\times10^{20}/cm^3$. The $\mu \propto n^{-1}$ tendency of single crystals is what is expected for the ionized dopant scattering[22]. According to an ionized dopant scattering model, $\mu$ can be dependent on $n$ through two different channels. One is the number of scattering centers, which makes $\mu \propto n^{-1}$, and the other is the strength of scattering, namely, the effect of the carrier density on the Debye length ($\lambda_D$), a screening length of an isolated charge. Since our single crystals showed metallic $\rho$ down to 2 K, they are in the degenerately doped regime so that $\lambda_D$ is already very small and probably changes very slowly. Therefore, the dominant effect on $\mu$ change as a function of $n$ is most likely to be the number of scattering centers, resulting roughly in the $\mu \propto n^{-1}$ relationship shown in Fig. 2(b). If we simply extrapolate our single crystal data, $\mu$ of the (Ba,La)SnO$_3$ system may reach a higher value with the decrease of $n$, although at lower doping level, other effects due to increased $\lambda_D$ and phonons may appear and limit its value. At the present stage, however, it was difficult to grow a single crystal with low La doping below 0.5 at% and homogeneous dopant distribution.

On the other hand, in our thin films, the decrease of $\mu$ has been observed when $n$ is decreased. Even though our films are dominantly epitaxial forms, there will be grain boundaries and dislocations, which act as strong charge traps and scattering centers for carriers reducing $n$ and $\mu$ simultaneously. The scattering centers are smeared out by the carrier-induced reduction of $\lambda_D$ resulting in the enhanced $\mu$ as increasing $n$. Our $\mu$ data of thin films follow closely the $\mu \propto n^{1/2}$ dependence. This dependence usually has been explained by dislocations in many other semiconducting films such as Ge[23] and GaN[24]. This may indicate that our films are mainly governed by dislocations. However, as the doping rate becomes higher than $4\times10^{20}$ cm$^{-3}$, the scattering by ionized dopants rather than dislocations or grain boundaries seems to start dominating. The main source for the dislocations and grain boundary in the thin films should be the SrTiO$_3$ substrate with an in-plane lattice constant (3.90 Å), which has a large lattice mismatch with BaSnO$_3$ (4.116 Å). Therefore, $\mu$ in thin films will be further improved when dislocation and grain boundary get reduced by, e.g., using a lattice-matched substrate.

To check the thermal stability of the (Ba,La)SnO$_3$ system, the resistance ($R$) of a Ba$_{0.96}$La$_{0.04}$SnO$_3$ thin film was investigated at elevated temperatures under Ar, O$_2$, as well as air atmosphere [Fig. 3]. Before monitoring $R$, we annealed our film in O$_2$ gas flow for 5 h at 700 °C to ensure that the film is in equilibrium with one atmosphere of oxygen pressure. We started monitoring $R$ of the film under an Ar gas flow to investigate the effect of the oxygen out-diffusion. While the temperature was maintained at 530 °C, $R$ decreased slowly, indicating that the oxygen started to diffuse out of the film in a relaxation time scale of a few hours. In the following annealing at 530 °C in O$_2$ gas, $R$ increased in a similar time scale, indicating that the oxygen diffused back into the film in a similar fashion. It is important to notice that the annealing at 530 °C in Ar gas resulted in $R$ change of only about 8% in 5 h, when the film reached a kind of equilibrium state. In the final annealing in air, the film showed a tiny decrease of $R$, only 1.7%, when it stayed at 530 °C for 5 h, demonstrating its exceptional stability in air.

The slight decrease of $R$ is mainly due to the reduced oxygen partial pressure in air. The entire oxygen out-diffusion and in-diffusion cycle was reproducible.

Only the 1.7% change of $R$ in air as observed in the 100-nm-thick $Ba_{0.96}La_{0.04}SnO_3$ film directly reflects the unusual stability of oxygen atoms inside the $SnO_2$ layers, as compared with other oxide materials. For example, ZnO films have shown about one order of $R$ change in air and more than four orders of magnitude change in $N_2$ when subject to similar annealing temperature and time conditions[25]. Based on the experimental results in Fig. 3, we can also estimate the amount of oxygen loss from the change of $R$. Since the initial conductance comes from the 4 at% La doping and each oxygen loss is equivalent to the two-La-ion doping based on its ionized charge value, the loss of oxygen is estimated to be only less than 0.2 at%. The oxygen diffusion constant at 530 °C turns out to be $(10^{-5} cm)^2/\pi^2 10^4 s = 10^{-15} cm^2 s^{-1}$, as estimated from the square of the film thickness divided by $\pi^2$ and its relaxation time[26]. This value is clearly lower than those of the well-known perovskite-based oxides such as titinates[27], cuprates[28] and manganites[26], the values of which are on the order of $10^{-8}$, $10^{-11}$, and $10^{-13}$ $cm^2 s^{-1}$ at a similar temperature region, respectively.

In conclusion, we showed that $(Ba,La)SnO_3$ materials have the unprecedentedly high $\mu$, only limited by ionized-dopant scattering in single crystals and by dislocations in epitaxial thin films. There is a possibility of the further improvement of $\mu$ by reducing the scattering both by dislocations and by the ionized dopants. The former can be achieved by developing lattice-matched substrates such as $BaSnO_3$, and the latter can be realized by introducing modulation doping as in the GaAs system. The multilayered two-dimensional structure necessary for modulation doping seems also plausible in these materials in light of its oxygen stability. Combined with its intrinsic transparency in a visible spectral range, high mobility, oxygen stability in elevated temperature, and three-dimensionality, the (Ba,La)SnO$_3$ material system offers great potential for transparent, high-frequency, high-power electronic devices.

**Acknowledgment(s)**

We thank Jaejun Yu, Jiyeon Kim, Tae Won Noh, Yun Daniel Park, and Tak Hee Lee for discussions. We acknowledge support from the NRF through the Accelerated Research Program (R17-2008-33-01000-0) and through the Creative Research Initiative (2010-0018300), and by MKE through the Fundamental R&D program for Core Technology of Materials.

**Figure captions**

**Figure 1** (**a**) BaSnO$_3$ has an almost ideal cubic perovskite with its lattice constant of 4.116 Å, and a SnO$_6$ octahedron inside the unit cell is plotted. (**b**) Optical microscope images of the flux grown BaSnO$_3$ (29.9 $\mu$m thick) and (Ba,La)SnO$_3$ (36.8 $\mu$m thick) single crystals, showing their optical transparency. (**c**) X-ray $\theta$-$2\theta$ scan results show the (00$l$) peaks in both single crystals and epitaxial thin films grown on SrTiO$_3$ (001) substrate. (**d**) The reciprocal space mapping of (103) peaks shows that the lattice constants are 4.127 Å in the growth direction and 4.107 Å on the plane, implying that there is a slight compressive strain on our 200-nm-thick film. $Q_X$ and $Q_Z$ are the in and out of plane components of a reciprocal space vector.

**Figure 2** (**a**) Resistivity $\rho$ and (**b**) mobility $\mu$ of the (Ba,La)SnO$_3$ thin films and single crystals are plotted as a function of carrier density $n$ at 300 K. The experimental data are shown as solid symbols while the dashed lines are guides to the eyes. The dotted lines represent $\mu \propto n^{-1}$ for the single crystal and $\mu \propto n^{1/2}$ for the thin films. Inset in Fig. 2**a** shows the temperature dependent resistivity of single crystal ($n = 1.62 \times 10^{20}$ cm$^{-3}$).

**Figure 3** (**a**) Temperatures and gas atmosphere were varied according to the profile in the upper panel and the resultant resistance variation is plotted in the lower panel. The film was 100 nm thick and temperature was maintained at 530 °C for 5 h. (**b**) The corresponding resistance vs. temperature is plotted. Resistance decreased (increased) by about 8 % under Ar (O$_2$) atmosphere in 5 h at 530 °C while it changed by only about 1.7 % in air.

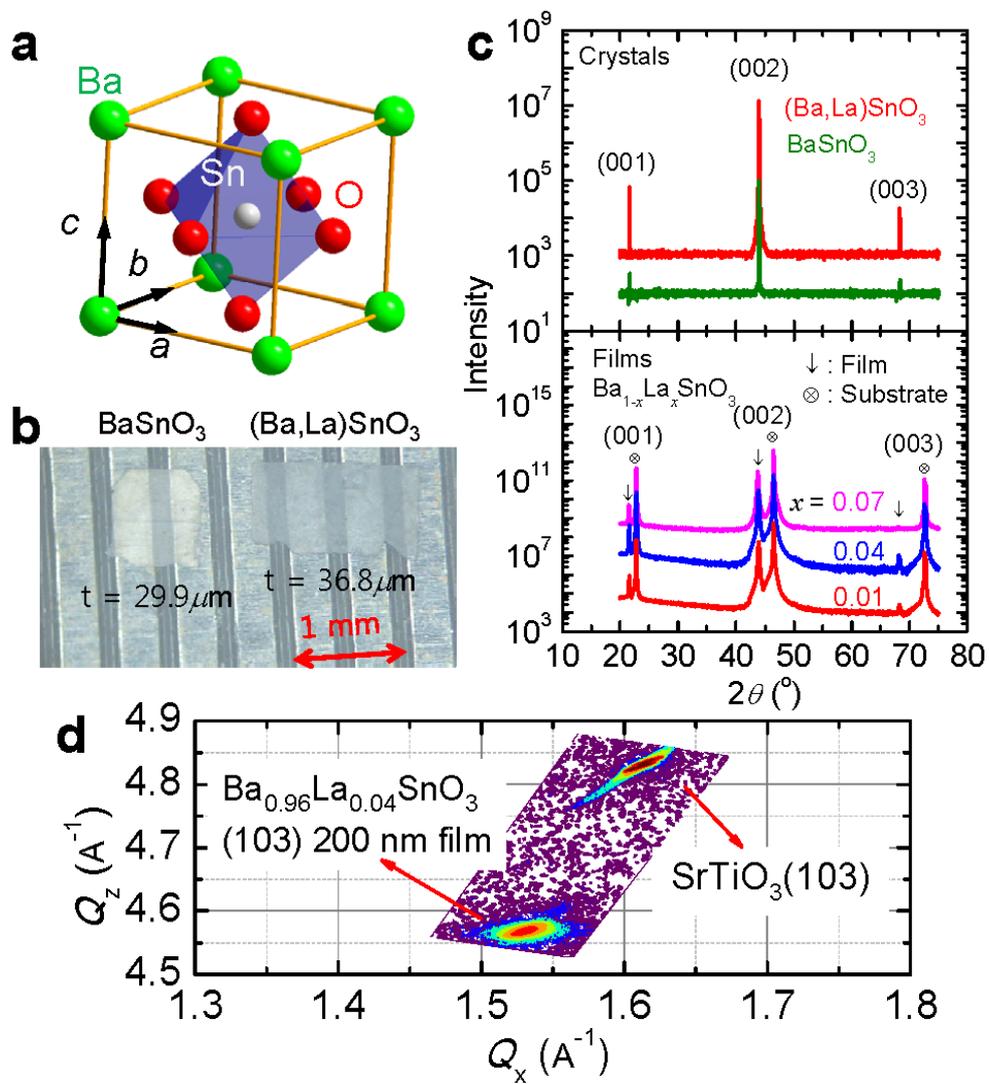

Figure 1. H. J. Kim *et al.*

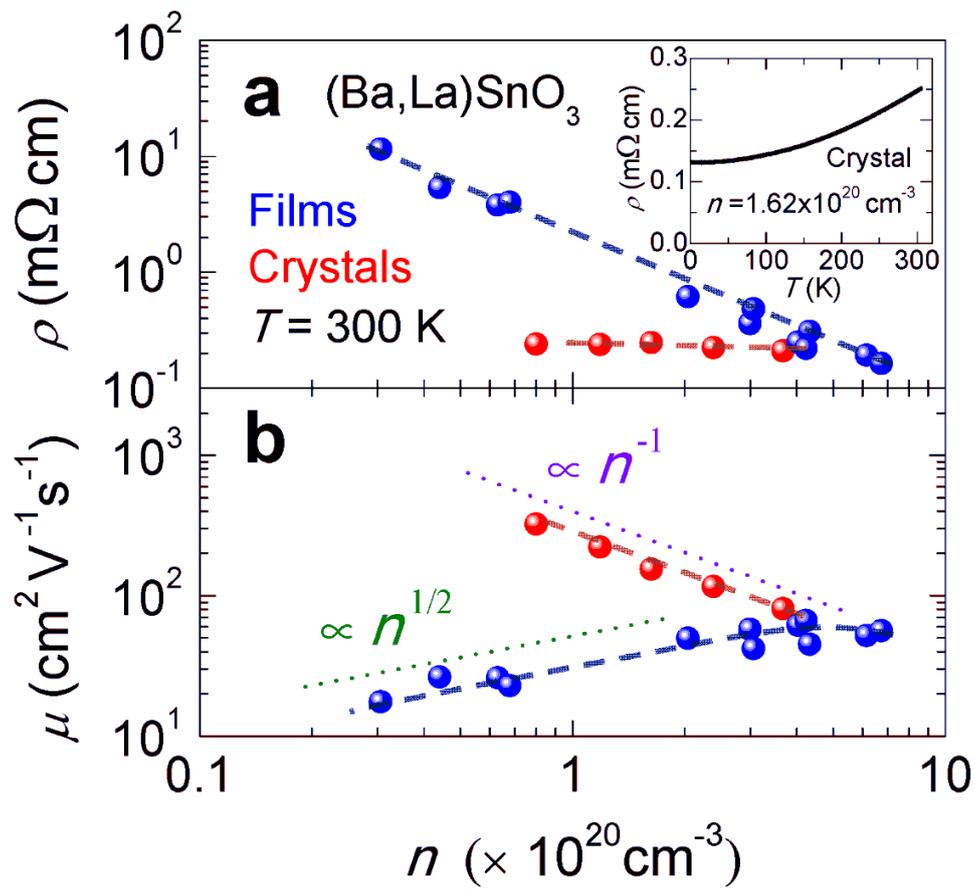

Figure 2. H. J. Kim *et al.*

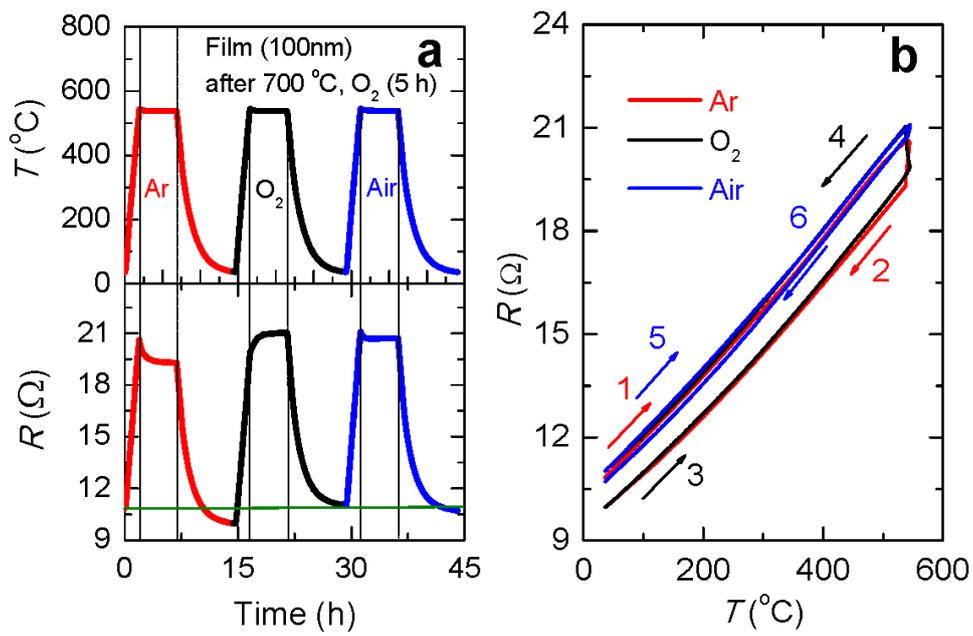

Figure 3. H. J. Kim *et al.*